\documentclass[preprint,3p,times]{elsarticle}

\usepackage{amssymb}
\usepackage{amsmath}

\usepackage[utf8]{inputenc}
\usepackage{url}
\usepackage{multirow}
\usepackage{capt-of} 
\usepackage{amsmath}
\usepackage{amssymb}
\usepackage{graphicx}
\usepackage{subfig}
\usepackage{parskip}
\usepackage{tikz}
\usepackage{verbatim}
\usepackage{rotating}
\usepackage{mathrsfs} 
\usepackage{float}
\usepackage{bm}
\usepackage{physics}
\usepackage{mathtools}
\usepackage{qcircuit}
\usepackage{array}
\usepackage{setspace}
\usepackage{ulem}
\usepackage{newfloat,algcompatible}
\usepackage[size=small]{caption}
\usepackage{etoolbox}
\usepackage{soul}
\AtEndEnvironment{algorithm}{\noindent\hrulefill\par\nobreak\vskip-8pt}

\DeclareFloatingEnvironment[
    fileext=loa,
    listname=List of Algorithms,
    name=\textbf{Algorithm},
    placement=tbhp,
]{algorithm}
\DeclareCaptionFormat{algorithms}{\vskip-15pt\hrulefill\par#1#2#3\vskip-6pt\hrulefill}
\algblock[Input]{Input}{EndInput}
\algblockdefx[Input]{Input}{EndInput}%
    [1]{\textbf{Input} #1}%
    {}

\algblock[Input]{Input}{EndInput}
\algblockdefx[Input]{Input}{EndInput}%
    [1]{\textbf{Input} #1}%
    {EndInput}

\algblock[For]{For}{EndFor}
\algblockdefx[For]{For}{EndFor}%
    [1]{\textbf{For} #1}%
    {EndFor}

\algblock[While]{While}{EndWhile}
\algblockdefx[While]{While}{EndWhile}%
    [1]{\textbf{While} #1}%
    {EndWhile}

\algblock[If]{If}{EndIf}
\algblockdefx[If]{If}{EndIf}%
    [1]{\textbf{If} #1}%
    {EndIf}

\usepackage[unicode=true,bookmarks=true,bookmarksnumbered=false,bookmarksopen=false,
 breaklinks=false,pdfborder={0 0 1},backref=false,colorlinks=true]
 {hyperref}
\hypersetup{
 linkcolor=magenta, urlcolor=blue, citecolor=blue, pdfstartview={FitH}, unicode=true}

\journal{Physica A: Statistical Mechanics and its Applications}

\begin{document}

\begin{frontmatter}



\title{Hybrid Real-Imaginary Time Evolution for Low-Depth Hamiltonian Simulation in Quantum Optimization}


\author[1]{Fei Li}
\ead{12031329@mail.sustech.edu.cn}
\affiliation[1]{organization={Department of Physics, Southern University of Science and Technology},
            city={Shenzhen},
            postcode={518055},
            state={Guangdong},
            country={China}}

\author[2]{Xiao-Wei Li}
\ead{lixiaowei@cdut.edu.cn}
\affiliation[2]{organization={Department of Physics, College of Physics, Chengdu University of Technology},
            city={Chengdu},
            postcode={610059},
            state={SiChuan},
            country={China}}


\date{\today}

\begin{abstract}
 Counterdiabatic (CD) driving is a powerful technique for accelerating adiabatic quantum computing. However, it becomes self-limiting in complex optimizations like the Sherrington-Kirkpatrick model: long evolution times $T$ needed to traverse crossings force the CD strength to scale as $1/T$, causing it to vanish before convergence and wasting the quantum resources invested in its implementation. We break this trade-off with a Hybrid adaptive variational quantum dynamics simulation (HAVQDS). HAVQDS combines adaptive real-time evolution for circuit compression with imaginary-time steps that suppress excitations at no extra gate cost. For the SK model (6--14 qubits), HAVQDS achieves higher approximation ratios than adiabatic or CD approaches, while reducing CNOT counts by 1--2 orders of magnitude, and avoids barren plateaus, ensuring non-vanishing parameter updates for scalable quantum optimization.
\end{abstract}



\begin{keyword}
Hybrid Real-Imaginary Time Evolution \sep Hamiltonian Simulation \sep Quantum Optimization 



\end{keyword}

\end{frontmatter}



\section{Introduction}
\label{sec:intro}

Adiabatic quantum computing (AQC)~\cite{RevModPhys.90.015002} provides a prominent framework for addressing combinatorial optimization problems~\cite{NPcomplete01,QuantumAnnealer17,farhi2000sat}, studying many-body models~\cite{Venturelli_2015}, and implementing algorithms such as the adiabatic Grover algorithm~\cite{Roland_2002} and adiabatic Deutsch-Jozsa algorithm~\cite{PhysRevLett.95.250503}.
In AQC, a quantum system is evolved from an easily preparable initial state to a final state encoding the solution to a problem. To maintain the system near the instantaneous ground state, the Hamiltonian must vary sufficiently slowly to avoid non-adiabatic transitions, often resulting in impractically long evolution times.

A powerful strategy to accelerate adiabatic protocols is the incorporation of a counterdiabatic (CD) term~\cite{RevModPhys.91.045001,PhysRevLett.123.090602,PhysRevApplied.15.024038}. This auxiliary Hamiltonian suppresses non-adiabatic transitions without altering the initial and final states~\cite{farhi2002cd,Zeng_2016,farhi2010quantumadiabaticalgorithmssmall}. Recent advances, including digitized-counterdiabatic quantum optimization~\cite{PhysRevResearch.4.L042030} and the digitized-counterdiabatic quantum approximate optimization algorithm~\cite{PhysRevResearch.4.013141}, have demonstrated that CD driving can significantly enhance optimization performance and success probability.

However, the practical utility of digitized CD methods is constrained by two inherent dilemmas. First, they introduce substantial circuit-depth overhead. Decomposing the time-evolution operator via Suzuki-Trotter methods results in a gate count that scales linearly with the total time $T$. For all-to-all connected models like the Sherrington-Kirkpatrick (SK) spin glass~\cite{SKmodel2012}, the per-step gate complexity is $O(n^2)$, leading to prohibitive resource requirements for large systems or long evolutions. Second, and more critically, a fundamental efficacy trade-off emerges: the strength of the CD Hamiltonian scales as $\dot{\lambda} \propto 1/T$. While a longer time $T$ is necessary to navigate narrowed gaps in complex energy landscapes, this very increase diminishes the strength of the CD term precisely when it is most needed. Consequently, the CD drive fails to suppress transitions before the computation converges, rendering the additional quantum resources invested in its implementation wasteful.

Several recent approaches have attempted to address these limitations from different perspectives. Counterdiabatic optimized local driving~\cite{PRXQuantum.4.010312} introduces a locally optimized control term within the counterdiabatic framework to dynamically manipulate the energy gap. However, its corrections remain tied to the evolution rate and may still require extensive classical optimization. In a different direction, feedback-based quantum algorithms inspired by CD~\cite{malla2024feedback} forgo the adiabatic theorem altogether and instead use measurement-based Lyapunov control to steer the system toward the ground state and therefore helps circumvent the CD strength constraint. On the circuit compression front, techniques such as those developed by Mc Keever and Lubasch~\cite{PRXQuantum.5.020362} use matrix product operators to compress adiabatic evolution circuits, achieving substantial reductions in gate count.

An alternative approach to mitigate circuit-depth growth is the adaptive variational quantum dynamics simulation (AVQDS)~\cite{PhysRevLett.130.040601}. As a variational quantum simulation method~\cite{Yuan_2019,Li_2017,Kokail_2019}, AVQDS dynamically selects and optimizes quantum gates to construct efficient circuits for simulating time evolution. Its adaptability makes it particularly suitable for noisy intermediate-scale quantum (NISQ) devices~\cite{Bharti_2022,Preskill_2018}. Nevertheless, AVQDS-based evolution under a purely adiabatic Hamiltonian still suffers from non-adiabatic transitions in systems with small energy gaps.

To overcome the limitations of both CD driving and standard variational adiabatic evolution, we propose a novel paradigm: a hybrid real-imaginary time adaptive variational quantum dynamics simulation (HAVQDS) algorithm. Our key insight is to replace the explicit CD term—which introduces algorithmic and resource overhead—with a hybrid-time evolution strategy. HAVQDS leverages AVQDS to perform efficient real-time evolution under the standard adiabatic Hamiltonian and interleaves this with \textit{variational} quantum imaginary-time evolution~\cite{Yuan2019theoryofvariational, mcardle2019variational} steps. Crucially, this variational implementation, which can be interpreted as natural gradient descent, provides an effective \textit{filtering} action that exponentially suppresses excited-state components. This process is executed by optimizing the parameters of the existing unitary parameterized quantum circuit, requiring no additional quantum gates, auxiliary systems, or non-unitary operations. This approach decouples the mechanism of non-adiabatic suppression from the real-time driving Hamiltonian, thereby bypassing the fundamental trade-offs associated with CD driving.

We demonstrate the efficacy of HAVQDS by applying it to the benchmark SK model. Numerical simulations involving up to 14 qubits show that our hybrid method achieves a higher final approximation ratio than conventional adiabatic (AD) and counterdiabatic (CD) digitized strategies.It accomplishes this while reducing the required number of CNOT gates by 1–2 orders of magnitude, establishing HAVQDS as a high-fidelity and resource-efficient solution for quantum optimization on near-term devices. Crucially, we demonstrate that HAVQDS inherently avoids barren plateaus—maintaining non-vanishing parameter updates across all simulated system sizes—and we provide a detailed analysis of the associated measurement costs, underscoring the practical feasibility of the algorithm for scalable quantum optimization.

\section{Dilemma in Counterdiabatic Acceleration}\label{sec:dilemma}
\begin{figure}[t]
    \centering
    \includegraphics[width=\linewidth]{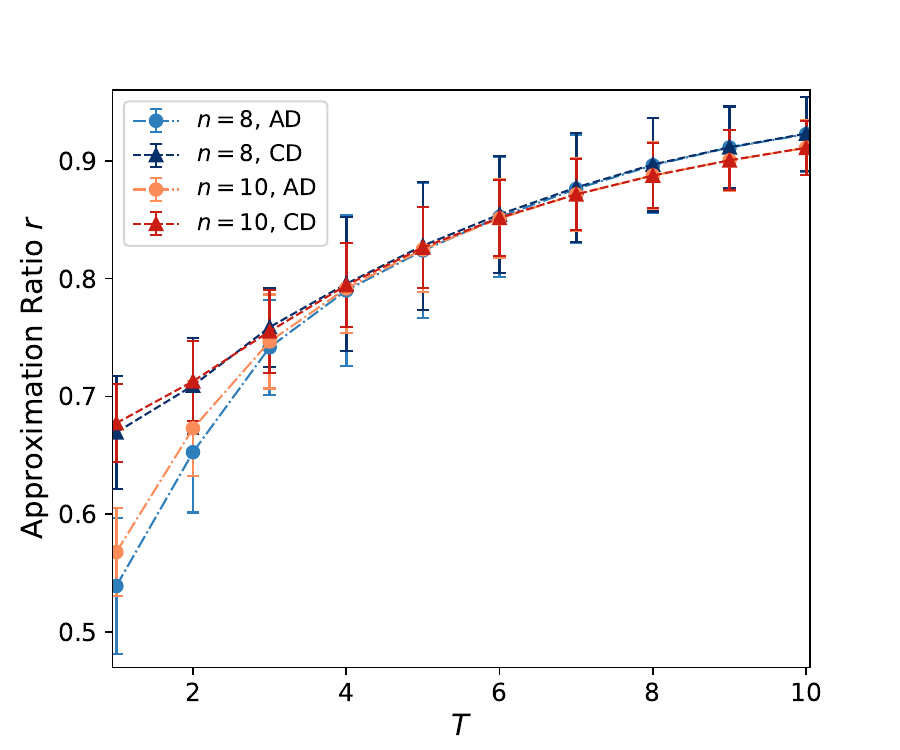}
    \caption{\textbf{Approximation Ratio $r$ vs Total Time $T$ for AD and CD Evolution Schemes.} The approximation ratio $r$ is plotted as a function of total time $T$ for AD and CD evolution strategies, for system sizes of $n=8$ and $n=10$ qubits. Data points represent the mean approximation ratio, with error bars indicating one standard deviation. The convergence of CD and AD performance at larger $T$ demonstrates the efficacy trade-off of the CD approach.}
    \label{fig:AC_CD_r}
\end{figure}

\begin{figure*}[t]
    \centering
    \includegraphics[width=\linewidth]{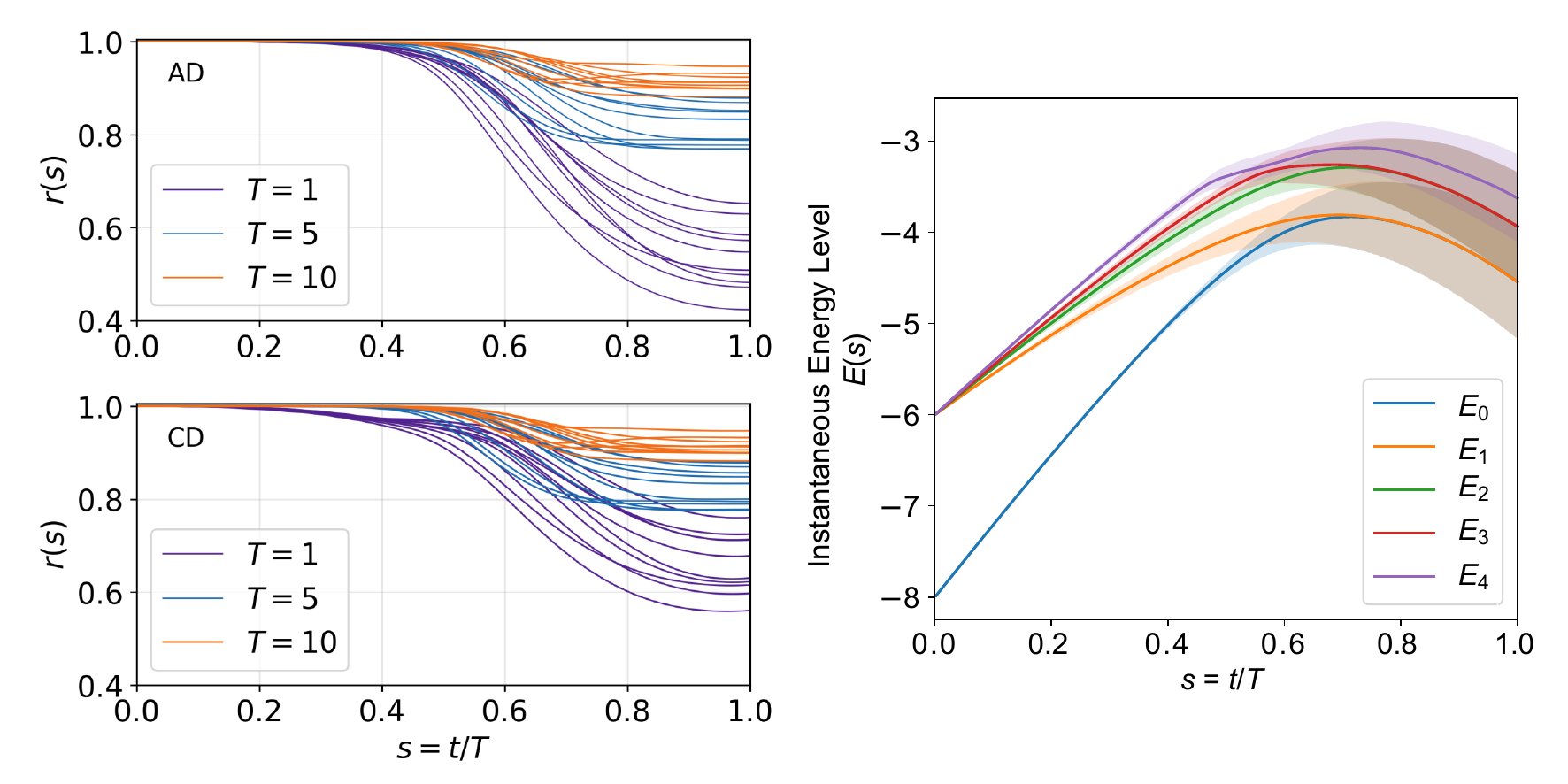}
    \caption{\textbf{Left panels:} Instantaneous approximation ratio $r(s)$ for AD (top) and CD (bottom) as a function of the dimensionless time $s=t/T$. Different colors correspond to annealing times $T=1, T=5$, and $T=10$ (purple, blue, orange). The decline in performance for $s > 0.4$ is evident for both protocols at longer $T$, highlighting the inability of the weakened CD term to prevent non-adiabatic transitions.
    \textbf{Right panel:} Instantaneous energy levels $E_0$ to $E_4$ (from bottom to top) of the Hamiltonian $H_{AD}(s)$ as a function of $s$. Shaded regions indicate standard deviation over 10 samples. Between $s=0.4$ and $s=0.8$, the energy gap narrows or even closes. All data in both panels are for an 8-qubit system.}
    \label{fig:dilemma}
\end{figure*}

To evaluate the performance of CD protocols in quantum optimization, we consider the SK model~\cite{SKmodel2012}, a canonical fully connected spin glass with all-to-all interactions. Its problem Hamiltonian is defined as:
\begin{equation}
    H_{SK}=-\sum_{i<j}^n J_{ij}\sigma_i^z\sigma_j^z,
\end{equation}
where the couplings $J_{ij}$ are drawn from a normal distribution with a mean of zero and a variance of $1/n$. The SK model exhibits a complex energy landscape featuring numerous avoided level crossings, making it a challenging and revealing benchmark for quantum annealing and digitized approaches.

Digitized counterdiabatic driving aims to accelerate adiabatic quantum computation by incorporating a CD term $H_{CD}$ into the original time-dependent Hamiltonian. This term suppresses non-adiabatic transitions, helping to maintain the system near the instantaneous ground state even under finite-time evolution. A Trotterized implementation facilitates the execution of such dynamics on digital quantum hardware. Further details on the CD formalism are provided in \ref{sec:CD}.

Despite its theoretical appeal, the practical application of digitized-CD is fundamentally limited by two dilemmas, starkly revealed in complex systems such as the SK model.

First, the \textbf{circuit depth scalability} poses a significant constraint. The gate count per Trotter step scales as $O(n^2)$ for all-to-all connected models such as SK, and the total circuit depth scales as $O(T \cdot n^2 / \delta t)$, where $\delta t$ is the time step. The inclusion of the CD term further increases the per-step gate count, resulting in prohibitive resource overhead for large systems or long annealing times $T$.

Second, and more critically, CD methods exhibit a \textbf{performance–efficacy trade-off}. The strength of the CD Hamiltonian is proportional to the rate of change of the scheduling parameter, $H_{CD} \propto |\dot{\lambda}| \propto 1/T$. This creates a paradox: although a longer total evolution time $T$ is necessary to traverse narrow avoided level crossings adiabatically, this same increase in $T$ weakens the CD term exactly when it is most needed. As shown in Fig.~\ref{fig:AC_CD_r} for system sizes $n=8$ and $n=10$, CD outperforms AD at short times ($T \lesssim 4$). However, as $T$ increases, the performance gap narrows and the curves converge, indicating that the advantage of CD diminishes with larger $T$. Consequently, in complex models such as the SK spin glass, the CD correction becomes ineffective near critical points involving energy level crossings, failing to suppress non-adiabatic transitions. This undermines the additional quantum resources invested in implementing CD, rendering it inefficient for challenging optimization problems.

To quantitatively illustrate these challenges, we simulate both Trotterized AD and Trotterized CD protocols for several 8-qubit and 10-qubit SK instances. Performance is evaluated using the instantaneous approximation ratio:
\begin{equation}
r(s) := \frac{E_{\text{max}}(s) - \langle H(s) \rangle}{E_{\text{max}}(s) - E_{\text{min}}(s)},
\end{equation}
where $E_{\text{max}}(s)$ and $E_{\text{min}}(s)$ are the instantaneous maximum and minimum eigenvalues of $H(s)$, respectively, and $s = t/T$ is the dimensionless time. The final approximation ratio is defined as $r = r(1)$.

Our results clearly demonstrate both dilemmas:
\textbf{1. Circuit Depth:} The number of CNOT gates scales linearly with $T$, with CD requiring approximately three times more gates than AD at the same $T$.
\textbf{2. Efficacy Loss:} As shown in the left panel of Fig.~\ref{fig:dilemma}, both AD and CD exhibit a sharp decline in the instantaneous approximation ratio $r(s)$ for $s \geq 0.4$. The right panel shows the corresponding energy spectrum, where the shaded region between $s = 0.4$ and $s = 0.8$ corresponds to a series of avoided crossings with significantly reduced energy gaps. It is in this critical region, where a strong CD drive is most needed, that its diminished strength ($\propto 1/T$) renders it ineffective, leading to rapid decreases in $r(s)$.

These findings highlight a fundamental limitation of conventional digitized-CD methods: they perform poorly precisely in the regimes where support is most needed, thus restricting their practicality for hard optimization problems on near-term quantum devices. This inherent dilemma motivates the development of a new paradigm that circumvents the need for an explicit CD term altogether. In the following section, we introduce our hybrid real-imaginary-time evolution approach (HAVQDS), which avoids these trade-offs by leveraging adaptive circuits and imaginary time filtering to suppress excitations, enabling high-fidelity evolution through complex energy landscapes without the resource overhead of CD driving.

\begin{algorithm}[t]
\caption{Hybrid Real-Imaginary Time Adaptive Variational Quantum Dynamics Simulation (HAVQDS)}
\label{alg:havqds}
\begin{algorithmic}[1]
\State \textbf{Input} $T, \delta t, \delta \tau, \Delta_{\text{cut}}, \epsilon_{\text{var}}, k_{\text{max}}$
\State $\ket{\psi} \gets \ket{+}^{\otimes n}$, $\bm{\theta} \gets \bm{0}$, $t \gets 0$ \Comment{Initialize}
\While{$t < T$}
    \State $s \gets t/T$
    \While{$\Delta > \Delta_{\text{cut}}$} \Comment{Adaptively expand ansatz}
        \State Expand ansatz with best operator from pool $\mathcal{P}$
        \State $\bm{\theta} \gets (\bm{\theta}, 0)$
    \EndWhile
    \State $\bm{\theta} \gets \bm{\theta} + A^{-1}\bm{C} \cdot \delta t$ \Comment{Real-time step}
    \State $t \gets t + \delta t$
    \If{$t \geq T$} \textbf{break} \EndIf

    \State $\sigma^2 \gets \langle H(t)^2 \rangle - \langle H(t) \rangle^2$ \Comment{Compute variance}
    \If{$\sigma^2 > \epsilon_{\text{var}}$} \Comment{Adaptive imaginary-time filtering}
        \State $j \gets 0$
        \While{$\sigma^2 > \epsilon_{\text{var}}$ \textbf{and} $j < k_{\text{max}}$}
            \State $\bm{\theta} \gets \bm{\theta} - (A^R)^{-1} C^R \cdot \delta \tau$ \Comment{Imaginary-time step}
            \State $\sigma^2 \gets \langle H(t)^2 \rangle - \langle H(t) \rangle^2$
            \State $j \gets j + 1$
        \EndWhile
    \EndIf
\EndWhile
\State \textbf{return} $\ket{\psi(\bm{\theta})}$, $\langle H(T)\rangle$
\end{algorithmic}
\end{algorithm}

\begin{figure*}[t]
    \centering
    \includegraphics[width=\linewidth]{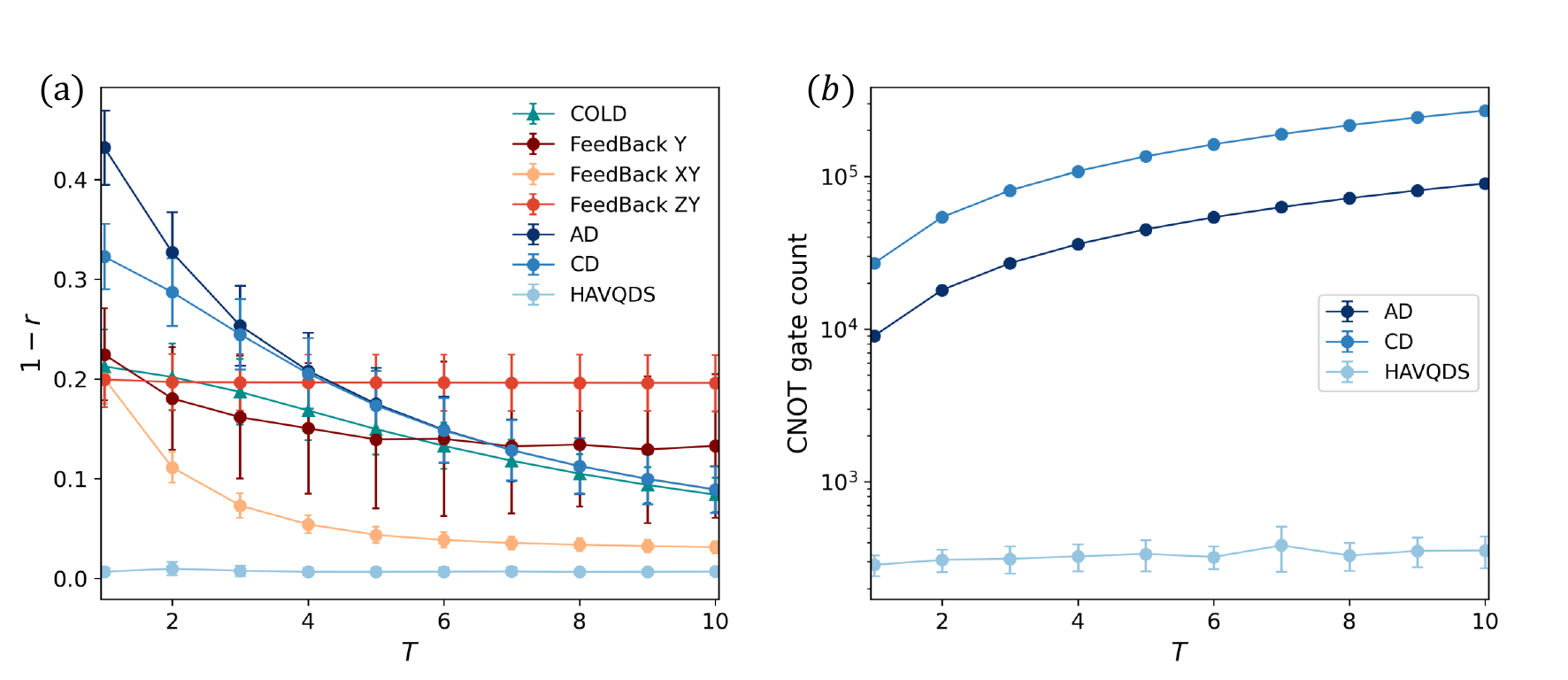}
    \caption{\textbf{Performance and efficiency of evolution schemes.} (a) Approximation error $1-r$ versus total time $T$. (b) CNOT gate count (log scale) versus $T$. Results are for a 10-qubit SK model, with error bars indicating standard deviation over multiple instances.}
    \label{fig:T_r_cnot}
\end{figure*}

\begin{figure*}[t]
    \centering
    \includegraphics[width=\linewidth]{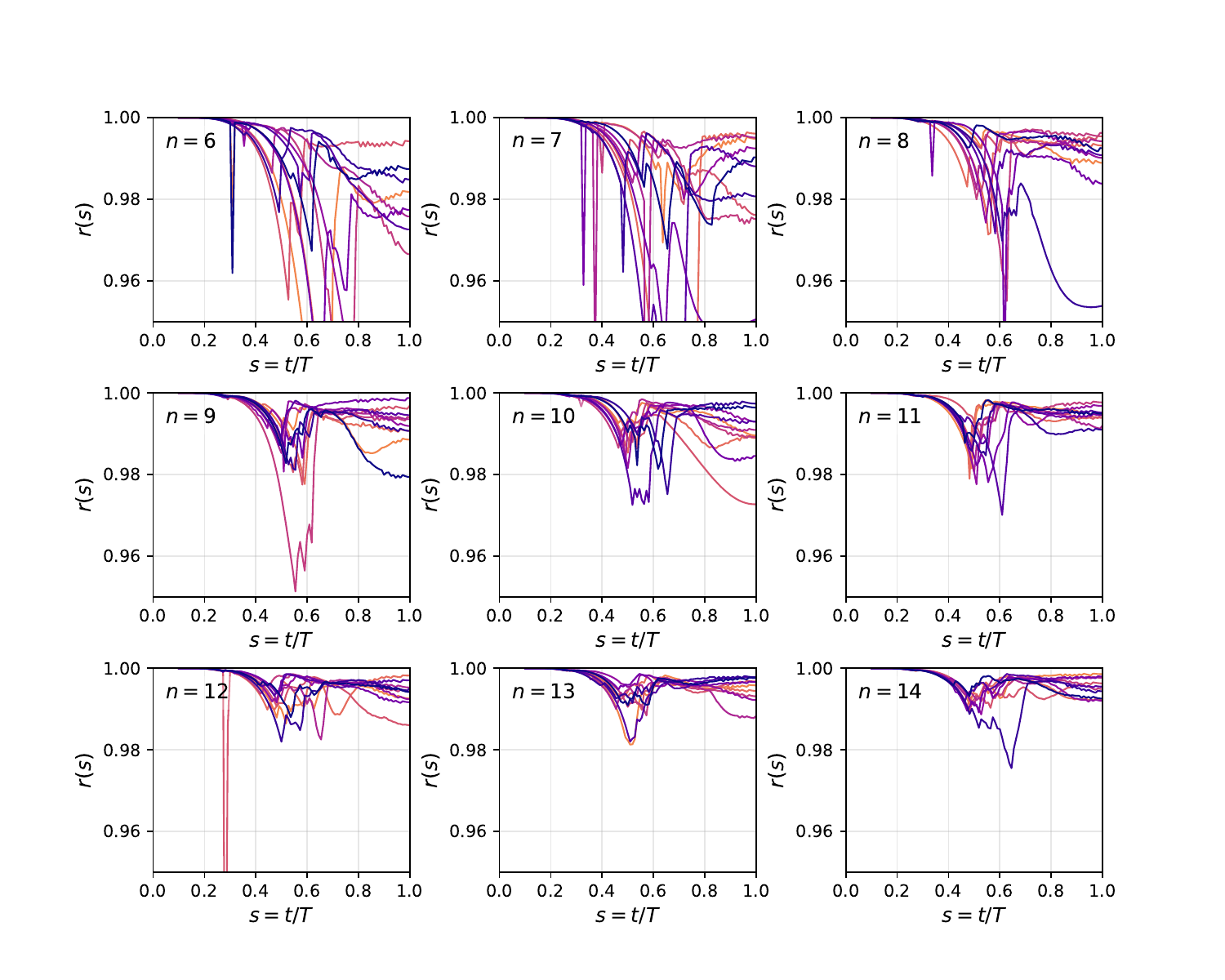}
    \caption{\textbf{Instantaneous approximation ratio across system sizes.} Each subplot corresponds to a different qubit number $n$ (6 to 14). The horizontal axis is the dimensionless time $s = t/T$ ($T=1$), and the vertical axis is the instantaneous approximation ratio $r(s)$. Each curve represents an independent instance of the SK model. HAVQDS maintains high fidelity throughout the evolution, especially in the region of avoided crossings ($s \approx 0.4$--$0.8$).}
    \label{fig:ratio_vs_n}
\end{figure*}

\begin{figure*}[t]
    \centering
    \includegraphics[width=\linewidth]{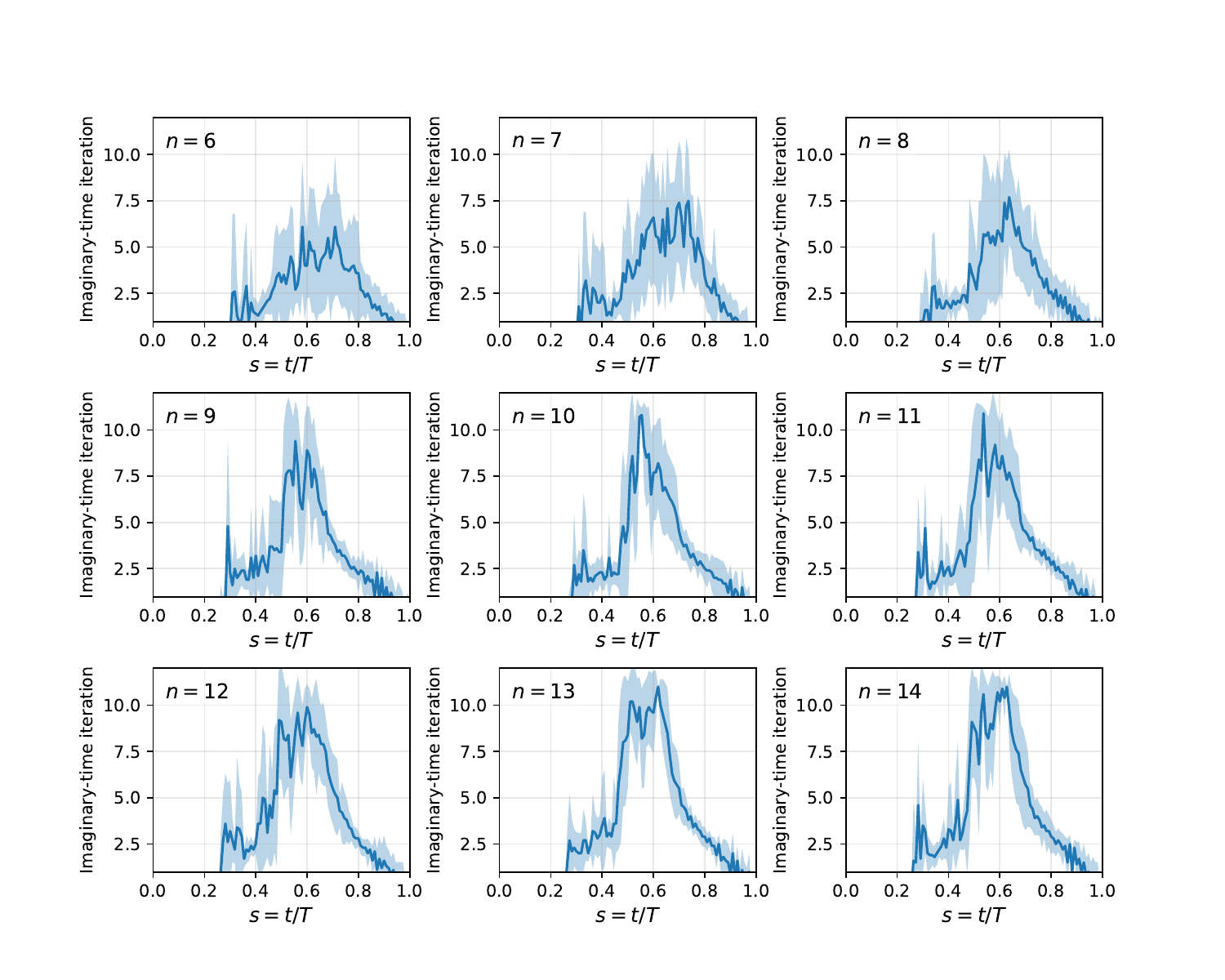}
    \caption{\textbf{Imaginary-time iterations versus rescaled time.} Each subplot shows the number of imaginary-time steps per real-time step as a function of $s = t/T$ for a system size $n$ (6 to 14). The shaded area represents the variance across 10 independent samples. The required number of steps peaks in the critical region of avoided crossings.}
    \label{fig:imag_vs_n}
\end{figure*}

\begin{figure}[t]
    \centering
    \includegraphics[width=0.8\linewidth]{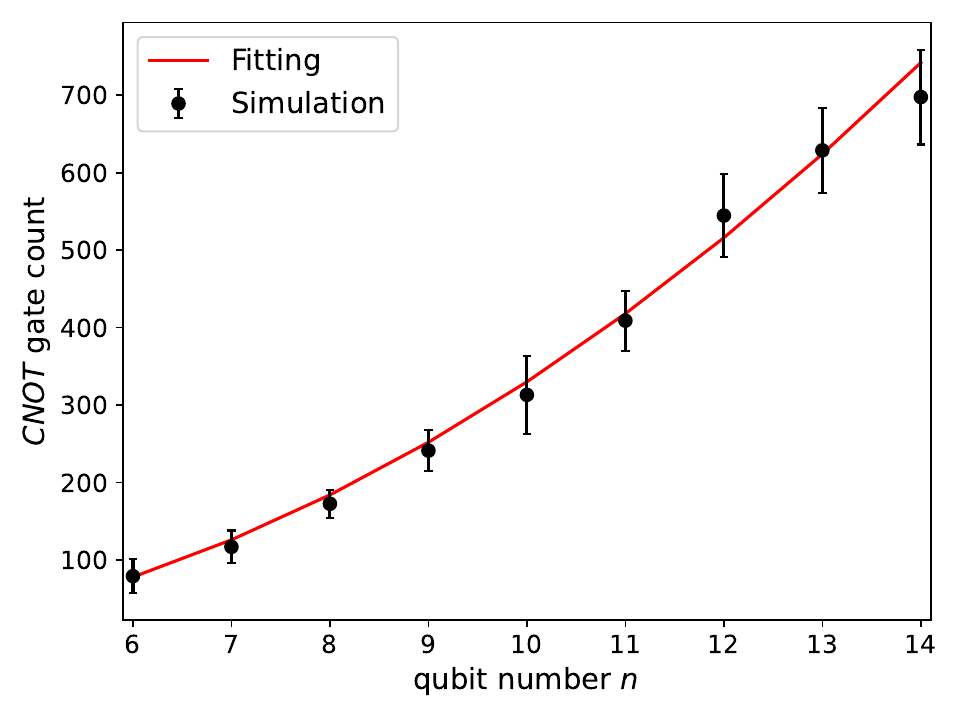}
    \caption{\textbf{CNOT gate count versus qubit number $n$.} The black dots show the mean total CNOT count for HAVQDS at $T=1$ for different system sizes (10 samples per size, error bars show standard deviation). The red line is a quadratic fit $5n^2 - 17n$, confirming the favorable scaling of the algorithm.}
    \label{fig:CNOT_count}
\end{figure}

\section{Hybrid Real-Imaginary Time Evolution Framework}\label{sec:hybrid}
The analyses in Sec.~\ref{sec:dilemma} reveal two fundamental bottlenecks in digitized counterdiabatic protocols: the prohibitive scaling of circuit depth and the self-limiting efficacy trade-off. We now introduce the HAVQDS algorithm, a new paradigm designed to simultaneously address both challenges. HAVQDS synergistically combines (i) the AVQDS framework to manage circuit complexity, with (ii) a hybrid-time evolution protocol that employs imaginary-time filtering to enhance ground-state convergence without adding gate overhead.

\subsection{Algorithm Design}
The HAVQDS algorithm is grounded in the variational quantum simulation framework. We consider a parameterized quantum state $\ket{\psi(\bm{\theta})}$, where the parameters $\bm{\theta}$ are now implicit functions of both real time $t$ and imaginary time $\tau$, i.e., $\bm{\theta} \equiv \bm{\theta}(t, \tau)$. The total state is thus defined as:
\begin{equation}
\ket{\psi(t, \tau)} := \prod_{\mu=1}^{N_{\theta}} e^{-i\theta_\mu(t,\tau) P_\mu}\ket{\psi_0}.
\end{equation}
This state is designed to approximate the solution of a hybrid-time evolution process. The goal is for $\ket{\psi(t, \tau)}$ to simultaneously satisfy, as closely as possible within the constraints of the variational ansatz, the real-time and imaginary-time Schr\"{o}dinger equations:
\begin{equation}
\begin{aligned}
i\frac{\partial }{\partial t} \ket{\psi(t,\tau)}&= H(t)\ket{\psi(t,\tau)}, \\
-\frac{\partial }{\partial \tau}\ket{\psi(t,\tau)} &= \left(H(t) - E(t,\tau)\right)\ket{\psi(t,\tau)},
\end{aligned}
\end{equation}
where $E(t,\tau) = \bra{\psi(t,\tau)}H(t)\ket{\psi(t,\tau)}$ is the instantaneous energy expectation value. The imaginary-time equation drives the state towards the ground state of $H(t)$ at a fixed $t$ by exponentially suppressing excited-state components.

The HAVQDS algorithm operationalizes this concept by alternating between evolution in $t$ and $\tau$. After initializing the system, it alternates between blocks of real-time evolution under the adiabatic Hamiltonian $H(s)$ and imaginary-time filtering steps applied to the instantaneous state. The real-time evolution is performed adaptively using the AVQDS method, which dynamically expands the ansatz to keep the McLachlan distance below a threshold $\Delta_{\text{cut}}$, thus ensuring accurate dynamics with near-minimal circuit depth; for details, see~\ref{app:avqds}. Crucially, the imaginary-time evolution is implemented variationally by solving Eq.~(\ref{eq:evolution}) for the parameter updates $\dot{\bm{\theta}}$, acting as a non-unitary filter that is executed \textit{unitarily} on the existing parameterized circuit. This requires no additional quantum gates, auxiliary systems, or post-selection, preserving the NISQ-compatibility of the entire protocol. The details are leveraged in~\ref{app:vqite}.

The adaptive ansatz expansion relies on a predefined operator pool $\mathcal{P}$. To ensure the expressibility of the variational ansatz for simulating the adiabatic Hamiltonian $H_{AD}(t)$ and effectively suppressing non-adiabatic transitions, we construct a comprehensive pool. It includes the Pauli terms constituting the problem Hamiltonian $H_{SK}$, the driver Hamiltonian $H_i = -\sum_i \sigma_i^x$, and importantly, the operators from the first-order CD term $H_{CD}^{(1)}$, which consists of $\sigma_i^y$ and $\sigma_i^z\sigma_j^y$ terms. This choice of pool is not arbitrary; it is justified by its proven completeness. The set $\{ \sigma_i^y, \sigma_i^z\sigma_j^y \}$ forms a universal pool for the Qubit-ADAPT-VQE framework~\cite{PRXQuantum.2.020310}, guaranteeing that any state in the Hilbert space can be reached, thereby ensuring the expressive power of our adaptive ansatz throughout the hybrid evolution.

A key innovation of HAVQDS is the adaptive execution of the imaginary-time evolution. Instead of applying a fixed number of steps, the filtering is triggered based on a physically motivated criterion: the variance of the instantaneous Hamiltonian, $\text{Var}(H(t)) = \langle H(t)^2 \rangle - \langle H(t) \rangle^2$. This variance serves as a proxy for non-adiabaticity; a large variance indicates significant population in excited states, deviating from the ideal adiabatic ground state. We set a threshold $\epsilon_{\text{var}}$. The imaginary-time evolution block is executed \textit{only if} $\text{Var}(H(t)) > \epsilon_{\text{var}}$, and continues until the variance falls below this threshold or a maximum number of steps $k_{\text{max}}$ is reached. This ensures that computational resources are spent only when necessary to correct the trajectory, further enhancing the efficiency of the overall algorithm.

The complete procedure is summarized in Algorithm~\ref{alg:havqds}.

\subsection{Mechanism of Excited-State Suppression}

The efficacy of the hybrid approach hinges on the ability of variational imaginary-time evolution to suppress excited-state components. Consider a quantum state at time $t$ expressed in the instantaneous energy eigenbasis of $H(t)$: $\ket{\psi(t)} = \sum_i \alpha_i \ket{\psi_i(t)}$. After applying an imaginary-time evolution step of duration $\tau$, the state becomes:
\begin{equation}
\ket{\psi(t, \tau)} = \frac{e^{-\tau H(t)} \ket{\psi(t)}}{\sqrt{\bra{\psi(t)} e^{-2\tau H(t)} \ket{\psi(t)}}}.
\end{equation}
The probability of being in the ground state, $p' = |\bra{\psi_0(t)}\ket{\psi(t, \tau)}|^2$, is given by:
\begin{equation}
p' = \frac{1}{1 + \sum_{i>0} \frac{|\alpha_i|^2}{|\alpha_0|^2} e^{-2\tau \Delta_{i0}}}\approx 1-\sum_{i>0}\frac{|\alpha_i|^2}{|\alpha_0|^2} e^{-2\tau \Delta_{i0}},
\end{equation}
where $\Delta_{i0} = E_i(t) - E_0(t)$ is the energy gap. In contrast, the probability without filtering is 
\begin{equation}
    p = \frac{1}{1 + \sum_{i>0} \frac{|\alpha_i|^2}{|\alpha_0|^2}}\approx 1- \sum_{i>0}\frac{|\alpha_i|^2}{|\alpha_0|^2}.
\end{equation} 
The imaginary-time step thus provides an \textit{exponential suppression} of excited-state components proportional to $e^{-2\tau \Delta_{i0}}$. This filtering action is achieved purely through optimization of the existing circuit parameters $\bm{\theta}$, introducing no new quantum gates.

\subsection{Numerical Results and Performance Analysis}

Numerical simulations on the SK model confirm that HAVQDS simultaneously achieves higher fidelity and lower resource costs compared to Trotterized alternatives. In our simulations, the step lengths for real and imaginary time evolution are set to $\delta t = 0.01$ and $\delta \tau = 0.05$, respectively. The McLachlan distance threshold $\Delta_{\text{cut}}$ is fixed at $0.05$, the variance threshold $\epsilon_{\text{var}}$ at $0.05$, and the maximum number of imaginary-time steps per block is set to $k_{\text{max}} = 11$. These parameters are found to strike a favorable balance between suppressing excited states and maintaining computational efficiency for the SK model instances studied. It is worth noting that the time-dependent Hamiltonian employed in our numerical simulations is the adiabatic Hamiltonian  \begin{equation}
    H_{AD}(t)=-[1-\lambda(t)]\sum_i^n \sigma_i^x+\lambda(t)H_{SK}.
\end{equation}
Existing counterdiabatic driving protocols and various other enhancements can be readily integrated with our approach, further extending its capability and performance.

To demonstrate the advantages of our method, we also compare it with the approaches proposed in Refs. \cite{PRXQuantum.4.010312,malla2024feedback}. Ref.~ \cite{PRXQuantum.4.010312} introduces counterdiabatic optimized local driving (COLD), which augments the standard CD framework with an additional local control term $\sum_i \sigma_i^z$. By optimizing the Fourier coefficients of this local drive, COLD can effectively manipulate the energy gap during evolution, mitigating the challenges posed by gap closure. In contrast, Ref.~\cite{malla2024feedback} presents a fundamentally different paradigm: a feedback-based quantum algorithm inspired by CD concepts. This approach completely abandons the guidance of the adiabatic theorem, instead leveraging quantum Lyapunov control theory to drive the system through measurement-based feedback, progressively lowering the energy expectation value. These two approaches represent distinct strategies—one refining the CD framework with optimized local driving, the other circumventing adiabatic constraints entirely through feedback control. Detailed descriptions of both schemes are provided in \ref{app:cold_feedback}.

\textbf{Approximation Ratio and Gate Count:} Figure~\ref{fig:T_r_cnot} compares HAVQDS against AD, CD, COLD, Feedback Y, Feedback XY and Feedback ZY for a 10-qubit SK model, demonstrating its superior performance and efficiency. HAVQDS consistently achieves a higher approximation ratio (i.e., lower $1-r$) across all evolution times $T$ (Fig.~\ref{fig:T_r_cnot}a), consistently outperforming all alternatives, while reducing CNOT gate counts by 1-2 orders of magnitude compared to AD and CD (Fig.~\ref{fig:T_r_cnot}b). By trotterization, the CNOT counts for COLD, Feedback XY, and Feedback ZY remain at levels comparable to that of CD, whereas the CNOT count for Feedback Y remains comparable to that of AD.
Critically, CD-based methods (CD, COLD, Feedback Y/XY/ZY) exhibit linear CNOT scaling with $T$ (requiring circuit compression techniques like Ref.~\cite{PRXQuantum.5.020362} for practical implementation), whereas HAVQDS, by design, replaces the CD term with hybrid real-imaginary time evolution and demonstrates saturation behavior without any compression overhead. This saturation arises because HAVQDS dynamically constructs an ansatz ``just expressive enough'' to represent the time-evolved state, decoupling resource requirements from $T$ (Fig.~\ref{fig:T_r_cnot}b). Consequently, HAVQDS achieves the same or better performance as CD-based methods without the need for external compression schemes.

The CNOT count for HAVQDS exhibits a crucial saturation behavior after an initial growth phase. This is a hallmark of the adaptive algorithm's efficiency. Unlike Trotterization, where circuit depth scales linearly with $T$ regardless of the underlying dynamics, HAVQDS dynamically constructs an ansatz that is just expressive enough to accurately represent the time-evolved state. Once this expressive, low-depth structure is discovered, the algorithm primarily optimizes within the existing parameter space rather than adding new gates. This indicates that the quantum resources (gate count, circuit depth) for simulating the adiabatic path are largely decoupled from the total time $T$, a significant advantage for long-time evolution and complex energy landscapes. 

\textbf{Scaling with System Size:} The performance advantage of HAVQDS is consistent across different problem scales. Fig.~\ref{fig:ratio_vs_n} shows the instantaneous approximation ratio $r(s)$ throughout the evolution for system sizes from 6 to 14 qubits. HAVQDS maintains a high $r(s)$ throughout the evolution, particularly in the critical region $s \in [0.4, 0.8]$ where avoided crossings occur. The number of imaginary-time steps required per real-time step, shown in Fig.~\ref{fig:imag_vs_n}, remains manageable and scales favorably with system size.  Finally, the total CNOT gate count for HAVQDS, shown in Fig.~\ref{fig:CNOT_count}, scales quadratically with the number of qubits $n$, adhering to the $O(n^2)$ scaling expected for the SK model and significantly outperforming the linear-in-$T$ scaling of Trotter methods.

In conclusion, the numerical results demonstrate that HAVQDS successfully breaks the trade-offs that limit CD protocols. By replacing the explicit CD term with an efficient variational imaginary-time filter, it achieves superior performance in complex energy landscapes while maintaining a low, scalable quantum resource footprint.

\section{Barren Plateaus and Measurement Cost}
\label{sec:classical_simulability}

\subsection{Barren Plateaus in Variational Quantum Algorithms vs. Variational Quantum Simulation}
The Barren Plateaus (BPs) phenomenon, originally identified in variational quantum algorithms (VQAs), refers to the exponential decay of gradient variances with increasing system size, which severely hinders parameter training~\cite{McClean_2018_11,cerezo2021Cost}. Traditional BP analysis assumes parameters are randomly initialized (e.g., uniformly or Haar-random) and optimized via unstructured exploration of the parameter space. Under this assumption, the gradient variance $\text{Var}[\nabla f(\theta)]$ serves as a natural metric: if it decays exponentially with qubit number, Chebyshev’s inequality implies the gradient is exponentially small with high probability, leading to optimization stagnation.

This probabilistic framework does not directly apply to variational quantum simulation (VQS), and in particular to our HAVQDS algorithm. In VQS, parameter evolution is governed by physical equations (the McLachlan variational principle) rather than random optimization. Parameters are not randomly initialized but start from a fixed initial state $|+\rangle^{\otimes n}$ and follow a deterministic trajectory constrained to a low-energy subspace. Thus, the relevant question is not ``what is the probability that a random parameter point lies in a flat region?'' but ``does the actual evolution trajectory ever enter a flat region?''.

To address this, we propose a direct monitoring approach: track the norm of the parameter update vector $\|\dot{\bm{\theta}}\|$ along the evolution path. If $\|\dot{\bm{\theta}}\|$ remains significantly above zero throughout, the algorithm does not suffer stagnation in practice. To provide a probabilistic assessment analogous to traditional BP analysis, we consider an ensemble of problem instances (different random SK models) and compute both the mean and variance of $\|\dot{\bm{\theta}}\|$ over that ensemble. A mean that remains size-independent and a variance that does not grow exponentially together imply that the updates are reliably non-vanishing, indicating that the algorithm avoids barren plateaus.

\subsection{Numerical Analysis of Update Dynamics}
We perform numerical simulations for system sizes $n = 5$ to $12$ qubits. For each size, we generate 100 random SK instances and run HAVQDS with total time $T=1$. We record $\dot{\bm{\theta}} = A^{-1}C$ at each time step and compute its norm.

\begin{figure}[H]
    \centering
    \includegraphics[width=\linewidth]{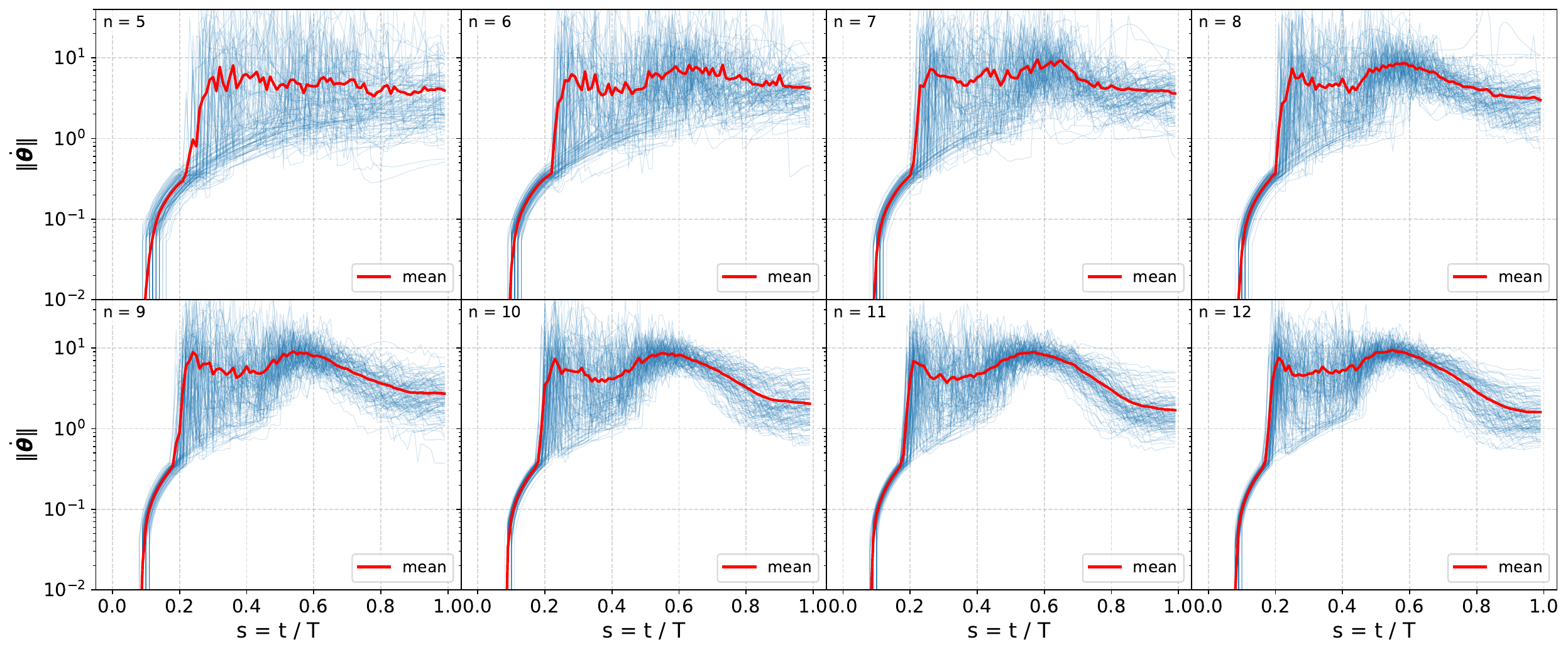}
    \caption{\textbf{Evolution of $\|\dot{\bm{\theta}}\|$ over $s=t/T$ for $n=5$ to $12$.} Each subplot corresponds to a specific $n$ value. The red lines represent the mean value, while the blue lines depict individual trajectories.}
    \label{fig:updatedynamics}
\end{figure}

\begin{figure}[H]
    \centering
    \includegraphics[width=\linewidth]{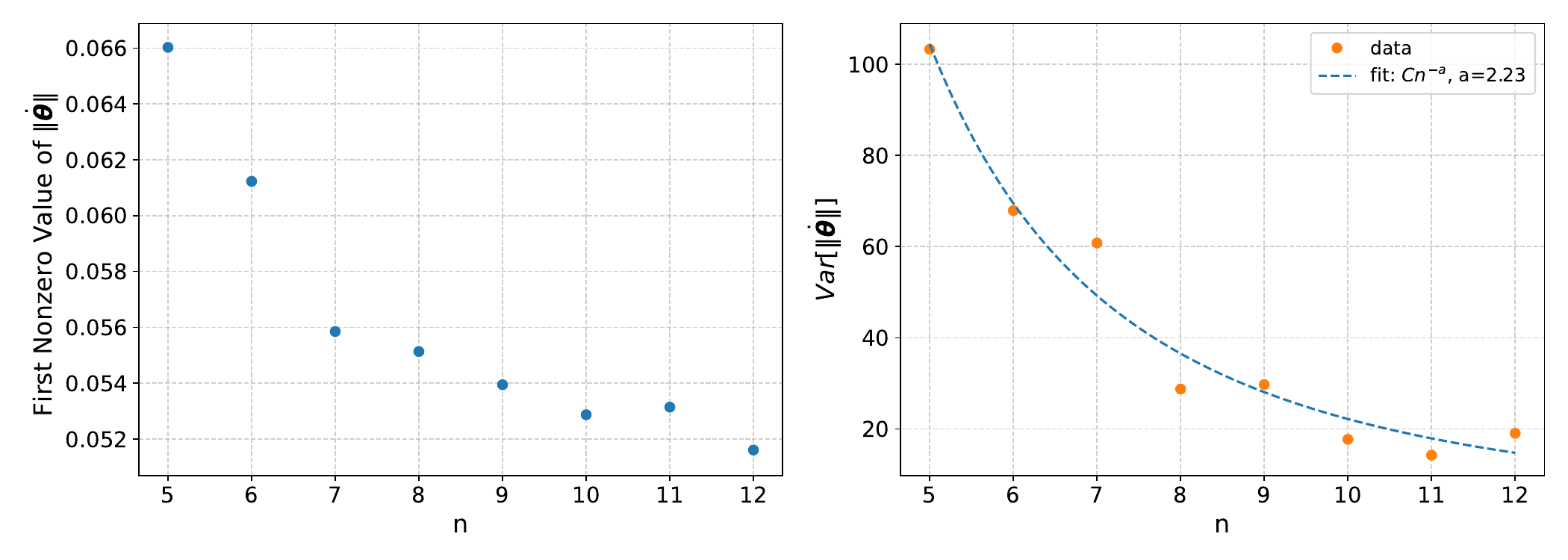}
    \caption{\textbf{Behavior of model parameter norms across different values of $n$.} \textit{Left panel}: Averaged over 100 samples per point (with negligible standard deviations), the results reveal the initial magnitude of the first nonzero parameter updates $\|\dot{\bm{\theta}}\|$ for each $n$. \textit{Right panel}: Orange dots represent empirical variance values, and the dashed line shows a power-law fit of the form $Cn^{-a}$ with an estimated exponent $a=2.23$. This illustrates that the variance of parameter norms decreases as $n$ increases, indicating increasing concentration of $\|\dot{\bm{\theta}}\|$ around its mean.}
    \label{fig:bp_simulation}
\end{figure}

\textbf{Results.} Fig.~\ref{fig:updatedynamics}  illustrates the evolution of the parameter update norm $\|\dot{\bm{\theta}}\|$ over dimensionless time $s = t/T$ for system sizes ranging from $n=5$ to $12$. For each $n$, we generated 100 random SK instances with $T=1$.
Three distinct dynamical regimes are observed across all system sizes. First, in the initial interval $(s\le0.1)$, the parameters remain effectively frozen ($\|\dot{\bm{\theta}}\|\approx0$); this corresponds to the regime where the McLachlan distance falls below the cutoff threshold $\Delta_{\mathrm{cut}}$, preventing updates. Second, around $s\approx 0.2$, the norm exhibits a sharp, monotonic increase as the system enters the active evolution phase.
Most critically, for $s>0.2$,  the mean norm stabilizes within the range of $10^{0}$ to $10^{1}$ and does not exhibit the exponential decay characteristic of Barren Plateaus. Although the variance among instances (indicated by the spread of blue lines) is visible, the mean trajectory remains robust across increasing $n$. This behavior confirms that our method maintains a non-vanishing gradient magnitude throughout the evolution, effectively avoiding the barren plateau region where gradients typically vanish exponentially with system size.

To further quantify this observation, we analyze the scaling of key statistics with 
$n$ in Fig.~\ref{fig:bp_simulation}. In traditional BP analysis, one typically examines the variance of gradient components, which equals the expectation value of squared gradient norm on the parameters ensemble when the gradient mean is zero. Here, we directly monitor the norm of the parameter update vector $\|\dot{\bm{\theta}}\|$ itself, which serves as a practical proxy for trainability. Fig.~\ref{fig:bp_simulation} shows that the first nonzero value of $\|\dot{\bm{\theta}}\|$ --- marking the onset of meaningful updates --- decreases only mildly with $n$, remaining above 
$0.05$ even at $n=12$. Importantly, the mean value of $\|\dot{\bm{\theta}}\|$ across the evolution remains of order $10^0$–$10^1$ for all system sizes (see Fig.~\ref{fig:updatedynamics}), indicating that updates do not vanish. Fig.~\ref{fig:bp_simulation} (right) further reveals that the variance of $\|\dot{\bm{\theta}}\|$ across the 100-instance ensemble scales polynomially as $Cn^{-2.23}$. While variance alone does not guarantee non-vanishing gradients, its polynomial decay—combined with the fact that the mean norm itself remains size-independent—provides strong numerical evidence that our adaptive filtering mechanism preserves trainability even as system size grows.

\textbf{Why HAVQDS avoids barren plateaus.} The evolution is constrained to a low-energy subspace, an exponentially small fraction of the full Hilbert space. In the Haar-random setting, most parameter points yield states with nearly identical Hamiltonian expectation values, giving vanishing gradients. Physical evolution keeps the system in the ``interesting'' part of parameter space where energy varies significantly. Additionally, imaginary-time steps filter the state back toward the ground-state manifold when non-adiabatic excitations occur, preserving update effectiveness.

\subsection{Measurement Cost}
\label{subsec:measurement_cost}

\textbf{Error propagation.} On real quantum devices, expectation values are estimated from a finite number of measurement shots, introducing statistical errors. Following the error-propagation analysis for variational time evolution \cite{PRXQuantum.2.030324}, the mean-squared error of $\dot{\bm{\theta}} = A^{-1}C$ satisfies
\begin{equation}
\mathbb{E}\bigl[\|\dot{\bm{\theta}}_{\text{shot}} - \dot{\bm{\theta}}\|^2\bigr] \leq \epsilon^2,
\end{equation}
where the accuracy $\epsilon$ is related to the total measurement count $N{\text{tot}}$ by
\begin{equation}
N_{\text{tot}} \leq \frac{K_A + K_C}{\epsilon^2},\qquad
K_A = 2\nu^4\text{Spc}[A^{-1}]^2\|C\|_\infty^2 f_A,\qquad
K_C = 2\nu^2\text{Spc}[A^{-1}]\text{Spc}[H(t)] f_C .
\end{equation}
Here $\nu$ denotes the number of parameters, and $\text{Spc}[A] = \|A\|_{F}^2 / d$ is the spectral mean (with $\|A\|_F$ the Frobenius norm and $d$ the dimension of $A$). The quantity $\text{Spc}[H_{AD}(\lambda)]$ for the Hamiltonian $H_{AD}(\lambda)$ can be expressed as
\begin{equation*}
\text{Spc}[H_{AD}(\lambda)] = \frac{\Tr [H_{AD}^\dagger H_{AD}]}{2^n}=\lambda^2\sum_{i<j}J_{ij}^2+(1-\lambda)^2n\approx \frac{(n-1)\lambda^2}{2}+n(1-\lambda)^2\sim O(n).
\end{equation*}
The factors $f_A, f_C \sim O(1)$ depend on the measurement grouping strategy.

\textbf{Regularization and measurement cost.} To enhance numerical stability and reduce the measurement overhead, we regularize $A^{-1}$ by truncating singular values below a threshold $\eta$. This reduces $\text{Spc}[A^{-1}]$, thereby lowering $K_A$, $K_C$, and the total required $N_{\text{tot}}$ for a given $\epsilon$. Specifically, since $\|A^{-1}\|^2 = \sum_i \sigma_i^2$, with $\sigma_i$ denoting the singular values of $A^{-1}$, we have
\begin{equation*}
\text{Spc}[A^{-1}] = \frac{\sum_i \sigma_i^2}{d} \leq \sigma_{\max}^2 \leq \eta^{-2}.
\end{equation*}
Fig.~\ref{fig:CNOT_count} suggests a scaling behavior of $\nu \sim n^2$; thus, the total measurement count can be estimated as
\begin{equation*}
N_\text{tot} \sim O\left(\frac{\max\left(n^8 \eta^{-4} \|C\|_\infty^2,  n^5 \eta^{-2}\right)}{\epsilon^2}\right).
\end{equation*}
While this bound may not be tight, this estimate highlights the substantial measurement cost that scales unfavorably with the system size, underscoring the need for more efficient approaches. Therefore, exploring a more structured operator pool to construct circuit ansatz -- thereby achieving a trade-off between the number of parameters and gate overhead -- and adopting metric-free variational quantum time evolution schemes \cite{PhysRevResearch.6.013143} to avoid the substantial overhead of directly measuring the matrix $A$ represents a crucial next step to mitigate this measurement overhead.

\section{Conclusion}

We have proposed a hybrid real-imaginary time evolution framework (HAVQDS) that integrates adaptive real-time dynamics with variational imaginary-time filtering to overcome the fundamental limitations of counterdiabatic protocols in quantum optimization.

Our approach directly addresses the two dilemmas of CD driving: it eliminates the need for an explicit CD term—and thus its associated circuit overhead and efficacy trade-off—while actively suppressing non-adiabatic transitions through an efficient, gate-free filtering mechanism. Numerical simulations on the SK model demonstrate that HAVQDS achieves a higher approximation ratio than both conventional adiabatic and counterdiabatic Trotterized approaches. Crucially, it accomplishes this with a drastic reduction in resource requirements, cutting CNOT gate counts by 1–2 orders of magnitude compared to CD.

By leveraging the adaptive flexibility of AVQDS and the exponential convergence properties of imaginary-time evolution, HAVQDS provides a robust alternative to CD strategies in systems with complex energy landscapes. Its ability to maintain high fidelity without increasing circuit depth makes it a highly promising algorithm for high-performance quantum optimization on NISQ devices. Future work will explore the application of HAVQDS to other challenging optimization problems and its experimental implementation on quantum hardware.

Importantly, we have shown that HAVQDS avoids the barren plateau problem that plagues many variational quantum algorithms. By monitoring the parameter update norm $\|\dot{\bm{\theta}}\|$ along the evolution, we find that its mean value remains of order $10^0$–$10^1$ across all system sizes, indicating that updates do not vanish as $n$ grows. Furthermore, the variance of $\|\dot{\bm{\theta}}\|$ across random instances scales polynomially with system size ($\sim n^{-2.23}$), providing additional evidence that the update strength remains concentrated and effective. This resilience stems from the physically guided, deterministic evolution constrained to low-energy subspaces. However, the measurement cost required to estimate the quantum geometric tensor remains a practical challenge, scaling unfavorably with $n$ under naive implementations. Future work will explore structured operator pools and metric-free variational time evolution schemes to mitigate this overhead, as well as extend HAVQDS to other optimization problems and experimental hardware platforms.

\section*{Acknowledgments}
This work was supported by the Foundation of Chengdu University of Technology.

%

\begin{appendix}
\numberwithin{equation}{section}

\section{Counterdiabatic quantum computation}~\label{sec:CD}
AD methods evolve the system via a time-dependent Hamiltonian $H_{AD}$, which is represented by
\begin{equation}
    H_{AD}(\lambda)=[1-\lambda(t)]H_i+\lambda(t)H_f,
\end{equation}
where $H_i=-\sum_i \sigma_i^x$ is the driven Hamiltonian and $H_f$ is the problem Hamiltonian.
They are connected by $\lambda(t)$, satisfying $\lambda(0)=0$ and $\lambda(T)=1$ for $t\in[0,T]$.

For the SK model, the problem Hamiltonian $H_f$ is given by 
\begin{equation}
    H_{SK}=-\sum_{i=1}^{n-1}\sum_{j>i}^n J_{ij}\sigma_i^z\sigma_j^z, 
\end{equation}
where the couplings $J_{ij}$ are independently normally distributed numbers with zero mean and  variance $1/n$. 

The main idea of counterdiabatic driving is to add an auxiliary term to the original Hamiltonian $H_{AD}$ and evolve the system
according to an effective Hamiltonian,
\begin{equation}\label{eq:cd}     H(\lambda)=H_{AD}(\lambda)+H_{CD}(\lambda).
\end{equation}
Here $H_{CD}$ is the CD term, which vanishes at the beginning and end of the protocol. To achieve this requirement, the mapping function $\lambda(t)$ can be selected as $\lambda(t)=\sin^2[\pi t/(2T)]$.

For the time-dependent Hamiltonian, the evolved state is given by
\begin{equation}
    \ket{\psi(T)} = \mathcal{T} \mathrm{exp}[-i\int^T_0 H(\lambda)dt]\ket{\psi(0)} ,
\end{equation}
where $\ket{\psi(0)}=\ket{+}^{\otimes n}$ with $\ket{+}=(\ket{0}+\ket{1})/\sqrt{2}$, and $\mathcal{T}$ is the time-ordering operator. 

As the paper shows~\cite{PhysRevApplied.15.024038,PhysRevLett.123.090602,PhysRevResearch.4.013141,PhysRevResearch.4.L042030,RevModPhys.91.045001}, the CD term is defined as $H_{CD}=\dot{\lambda}A_\lambda$.
Here $A_\lambda$ is known as the adiabatic gauge potential responsible for the non-adiabatic transitions.
Although $A_\lambda$ in the exact form is theoretically optimal, in practice it is often more feasible and efficient to choose CD terms in the approximate form. 
The approximate form is not only computationally simpler, but also experimentally easier to implement, and still can effectively accelerate the adiabatic process~\cite{PhysRevLett.123.090602,PhysRevApplied.15.024038,PhysRevResearch.4.013141,PhysRevResearch.4.L042030}.
We consider a general way to choose the ansatz using the nested commutator approach of the adiabatic gauge potential.
\begin{equation}
     A^{(l)}_\lambda =i\sum^l_{k=1}\alpha_k(t)\underbrace{[H_{AD},[H_{AD},\dots[H_{AD}}_{2k-1},\partial_\lambda H_{AD}]]].
\end{equation}
When $l\in \infty$, we will get the exact gauge potential. 
Here $\alpha_k(t)$ is the CD coefficient, obtained by minimizing the operator distance between the exact gauge potential and the approximate gauge potential, 
which is equivalent to minimizing the action,
\begin{equation}
    S_{\lambda}(A_\lambda)=\mathrm{Tr} [(\partial_\lambda H_{AD}+i[A_\lambda, H_{AD}])^2]
\end{equation}

Consider the specific form of the corresponding time-dependent Hamiltonian 
\begin{equation}
    H_{AD}(\lambda) =\lambda(t)[-\sum_{i<j}J_{ij}\sigma^z_i\sigma^z_j+ \sum_i h_i\sigma^z_i]-[1-\lambda(t)][\sum_i\sigma_i^x],
\end{equation} 
and the general two-local CD term 
\begin{equation}
    H_{CD}(\lambda) =\sum_i\alpha_i(\lambda)\sigma^y_i+ \sum_{i\neq j}\beta_{ij}(\lambda)\sigma^z_i\sigma^y_j+ \gamma_{ij}(\lambda)\sigma^x_i\sigma^y_j,
\end{equation} 
where $\sigma^x$, $\sigma^y$ and $\sigma^z$ are the Pauli operators, and the CD coefficients $\alpha_i, \beta_{ij}$, and $\gamma_{ij}$ are obtained by variational minimization. $J_{ij}$ is coupling and $h_i$ is the vertical field.
Finally, we approximate the CD term using first-order nested commutator~\cite{PhysRevResearch.4.L042030}
\begin{equation}
     H^{(1)}_{CD} (\lambda)=-2\dot{\lambda}\alpha_1(t)[\sum_i h_i\sigma^y_i-\sum_{i<j}J_{ij}(\sigma^y_i \sigma^z_j+\sigma^z_i\sigma^y_j)].
\end{equation}
Here, the CD coefficient 
\begin{equation}
    \alpha_1(t)=-\frac{1}{4}[\sum_i h^2_i+2\sum_{i<j}J^2_{ij}]/R(t),
\end{equation}
where $R(t)$ is given by
\begin{equation}
\label{footnote}
\begin{split}
    R(t)=&[1-2\lambda(t)]\left[\sum_i h^2_i+8\sum_{i<j} J^2_{ij}\right]
    +\lambda(t)^2\left[\sum_i h^2_i+\sum_i h^4_i +8\sum_{i<j}J^2_{ij}+2\sum_{i<j}J^4_{ij}+6\sum_{i=j}h^2_i J^2_{ij}+6\sum_{i<j}\sum_{k<l}J^2_{ij}J^2_{kl}\right]. 
\end{split}
\end{equation}

\section{Setup of the Simulation for Counterdiabatic Optimized Local Driving and Feedback-Based Counterdiabatic Driving }~\label{app:cold_feedback}
The COLD method is based on continuous-time evolution, where an additional local driving term is incorporated into the adiabatic Hamiltonian, written as
\begin{equation}
H_{\bm{\beta}}(\lambda)=H_{AD}(\lambda)+f(\bm{\beta},\lambda)\sum_{i=1}^n\sigma_i^z,
\end{equation}
with $\lambda=\sin^2[\pi t/(2T)]$ being the time parameterization adopted in our work, and
\begin{equation}
f(\bm{\beta},\lambda)=\sum_{j=1}^k \beta_j \sin(2k\pi t/T),
\end{equation}
where $\beta_j$ are the Fourier coefficients to be optimized. In our simulation, we set $k=3$. Based on this, the counterdiabatic Hamiltonian for $H_{\beta}(t)$ is constructed as
\begin{equation}
H_{CD}=H_{\beta}+\dot{\lambda} A_\lambda.
\end{equation}
Here, the adiabatic gauge potential under the first-order nearest-neighbor commutator approximation takes the form
\begin{equation}
A_\lambda = i\alpha(\lambda)[H_{\beta},\partial_\lambda H_{\beta}],
\end{equation}
and $\alpha(\lambda)$ is determined by solving the following minimization problem:
\begin{equation}
\begin{aligned}
\alpha(\lambda)&:=\arg\min_{\alpha}\Tr\left\{ \left(\partial_\lambda H_\beta + i[A_\lambda, H_\beta]\right)^2 \right\}\\
& = \frac{2\sum_{i<j}J_{ij}^2+n\gamma^2}{R(\lambda)},
\end{aligned}
\end{equation}
where $\gamma=f+(1-f)\partial_{\lambda}f$, and
\begin{equation}
    R(t)=-\frac{1}{4}\left\{\sum_i(2\lambda\sum_{j>i}J_{ij}^2+\gamma f)^2+n\gamma^2(1-\lambda)^2+2\left[4(1-\lambda)^2+(f+\gamma\lambda)^2\right]\sum_{i<j}J_{ij}^2+2\lambda^2\sum_{i}\left[(\sum_{j>i}J_{ij}^2)^2-2\sum_{j>i}J_{ij}^4\right]\right\}.
\end{equation}
Setting the initial state $\ket{\psi(0)}=\ket{+}^{\otimes n}$, we simulate the time evolution using a Trotterization with step size $\delta t = 0.01$, and take the expectation value of $H_{SK}$ as the loss function $Loss(\bm{\beta})$ for the optimization of the 3-order Fourier coefficients $\bm{\beta}=(\beta_1,\beta_2,\beta_3)^T$. The optimization is performed using the \textit{Powell} algorithm, which is suggested by the literature.

Feedback-based counterdiabatic driving does not rely on the adiabatic theorem; instead, it leverages measurement feedback guaranteed by quantum Lyapunov control theory to lower the energy expectation value. The control equation is given by
\begin{equation}
i\frac{d}{dt}\ket{\psi(t)}=\left[H_{SK}+\beta(t)\sum_{i=1}^n\sigma_i^z+\gamma(t)H_{CD}\right]\ket{\psi(t)}.
\end{equation}
Unlike the COLD approach, this method does not require an explicit detailed construction of the CD term. The original work provides several choices for $H_{CD}$, among which we test the two that exhibit better performance:
\begin{itemize}
\item \textbf{FeedBack Y}: $H_{CD}^{Y}:=\sum_{i=1}^n \sigma_i^y$,
\item \textbf{FeedBack ZY}: $H_{CD}^{ZY}:=-\sum_{i<j}^n J_{ij} (\sigma_i^z\sigma_j^y+\sigma_i^y\sigma_j^z)$.
\item \textbf{FeedBack XY}: $H_{CD}^{XY}:=-\sum_{i<j}^n J_{ij}(\sigma_i^x\sigma_j^y+\sigma_i^y\sigma_j^x)$.
\end{itemize}
The parameters are updated according to the following rules:
\begin{equation}
\begin{aligned}
\beta^{(k+1)}&=\frac{i\alpha}{n}\bra{\psi(t_k)}[H_{SK},\sum_{i=1}^n\sigma_i^z]\ket{\psi(t_k)},\\
\gamma^{(k+1)}&=\frac{i\alpha}{n}\bra{\psi(t_k)}[H_{SK},H_{CD}]\ket{\psi(t_k)},
\end{aligned}
\end{equation}
where $t_k=k\delta t$ with $\delta t=0.01$, and we set the coefficient $\alpha=3$.

\section{Adaptive Variational Quantum Dynamical Simulation Framework}\label{app:avqds}
Hamiltonian simulation is a core task in quantum computing, simulating the evolution of physical systems over time.
Traditional simulation methods usually require many quantum gate operations, which results in a large circuit depth, increasing the execution time and error rate.
Ref.~\cite{PhysRevLett.130.040601, yao2021adaptive} presents an adaptive product formula method for efficiently simulating Hamiltonian evolution in quantum systems. 
By dynamically adjusting the Hamiltonian decomposition, this method significantly reduces the number of quantum gates and circuit depth required for simulation, thus improving the accuracy and practicality of simulation.

Herein, we follow the formula framework of variational quantum dynamics simulation proposed by Ref.~\cite{yao2021adaptive}, which employs the von Neumann equation 
\begin{equation}
    \frac{d\rho}{dt}+\mathcal{L}(\rho)=0,
\end{equation}
with $\rho=\ket{\psi(t)}\bra{\psi(t)}$ and $\mathcal{L}=-i[H,\rho]$, to determines the dynamic processes.
In the variational quantum simulation, the time-dependent state $\ket{\psi(t)}$ is represented by an approximated time-evolved state
\begin{equation}
    \ket{\psi(\bm{\theta}(t))}:=\prod_{\mu=1}^{N_{\bm{\theta}}}e^{-i \theta_\mu P_\mu }\ket{\psi_0},
\end{equation}
where the parameters $\bm{\theta}(t):=(\theta_1(t),\cdots, \theta_{N_{\bm{\theta}}}(t))$ are time-dependent, and $\mathcal{P}:=\{P_{\mu}\}$ is a set of Hermitian operators.
The time evolution with time step $\delta t$ can be determined by 
\begin{equation}
    \bm{\theta}(t+\delta t)=\bm{\theta}(t)+\delta t\dot{\bm{\theta}}.
\end{equation}

Consequently, the key step in realizing the time evolution is to obtain the time derivative of parameters $\dot{\bm{\theta}}$. This can be approached by McLachlan's variational principle, which aims to minimize the square McLachlan distance
\begin{equation}
\begin{aligned}
    \Delta^2 &=\left|\left|\frac{d\rho}{dt}+\mathcal{L}(\rho)\right|\right|^2\\
    &=\dot{\bm{\theta}}^\top A\dot{\bm{\theta}}-2\dot{\bm{\theta}}\cdot \bm{C}+2\text{Var}_{\bm{\theta}}(H),
\end{aligned}
\end{equation}
where $||\sigma||=\sqrt{\Tr(\sigma^\dagger\sigma)}$ is the Frobenius norm of the matrix $\sigma$, $\text{Var}_{\bm{\theta}}(H)=\langle H^2\rangle_{\bm{\theta}}-\langle H\rangle^2_{\bm{\theta}}$ with $\langle H\rangle_{\bm{\theta}}=\bra{\psi(\bm{\theta})} H\ket{\psi(\bm{\theta})}$. The matrix $A$ is defined as 
\begin{equation}
\begin{aligned}
    A_{\mu\nu}:=&2\text{Re}[
    \bra{\partial_\mu \psi(\bm{\theta})}\partial_\nu\psi(\bm{\theta})\rangle \\&+\bra{\partial_\mu\psi(\bm{\theta})}\psi(\bm{\theta})\rangle\bra{\partial_\nu\psi(\bm{\theta})}\psi(\bm{\theta})\rangle
    ],
\end{aligned}
\end{equation}
where $\partial_\mu=\frac{\partial \ket{\psi(\bm{\theta})}}{\partial \theta_\mu}$.
The vector $\bm{C}$ is defined as 
\begin{equation}
\begin{aligned}
    \bm{C}_{\mu}:=&2\text{Im}[\bra{\partial_\mu\psi(\bm{\theta})}H\ket{\psi(\bm{\theta})}\\
    &+\bra{\psi(\bm{\theta})}\partial_\mu\psi(\bm{\theta})\rangle\langle H\rangle_{\bm{\theta}}].
\end{aligned}
\end{equation}
As mentioned in Ref.~\cite{Yuan_2019}, the square McLachlan is a metric for the accuracy of quantum dynamical evolution.

The minimization of $\Delta^2$ with respect to the parameters $\dot{\bm{\theta}}$ yields the equation of motion 
\begin{equation}\label{eq:update}
    A\dot{\bm{\theta}} = \bm{C},
\end{equation}
Thus, the time derivative of parameters is given by $\dot{\bm{\theta}}=A^{-1}\bm{C}$.

Next, we briefly review the adaptive variational quantum dynamics simulation proposed in Ref.~\cite{PhysRevLett.130.040601,yao2021adaptive}. 
In adaptive variational quantum dynamics simulation, the approximated time-evolved state is constructed by an adaptive product formula method, in which the set of Hermitian operators $\{P_\mu\}$ is expanded by incorporating additional operators from a predefined pool, ensuring the McLachlan distance $\Delta$ remains below a specified threshold $\Delta_{\text{cut}}$. 

Without loss of generality, we consider the following time-dependent Hamiltonian:
\begin{equation}
H(t) = \sum_{j=1}^L h_j(t) P_j, \qquad P_j \in \mathcal{P},
\end{equation}
which drives the dynamic process. We choose the set \(\mathcal{P}\) as the operator pool. 

At \(t = 0\), we prepare the initial state \(\rho_0 = (\ket{+}\bra{+})^{\otimes n}\); For each time step \(t\), we select operators from the operator pool \(\mathcal{P}\) to augment the set of operators \(\{P_\mu\}\) in the approximate time-evolved state and compute the McLachlan distance \(\Delta\), until \(\Delta \leq \Delta_{\text{cut}}\); Based on the updated set of operators \(\{P_\mu\}\), we extend the corresponding parameter vector \(\bm{\theta}^{(t)}\) by appending new zero entries, \(\bm{\theta}^{(t)} \oplus \bm{0}_{\text{new}}\), and then calculate \(\dot{\bm{\theta}}\) using Eq.~\eqref{eq:update}; The variational parameters at next time step \(t + 1\) are updated as \(\bm{\theta}^{(t+1)} = \bm{\theta}^{(t)} \oplus \bm{0}_{\text{new}} + \delta t \dot{\bm{\theta}}\). We repeat these steps iteratively until reaching the final time \(T\).

\section{Variational quantum imaginary-time evolution}\label{app:vqite}

Quantum Imaginary-Time Evolution (QITE) is a promising method to prepare the ground state of quantum systems on near-term quantum computers~\cite {motta2020determining}. It enables faster buildup of quantum correlations compared to real-time evolution, which is constrained by the Lieb-Robinson bond~\cite{bench2019making}, and it is guaranteed to converge to the ground state under ideal conditions~\cite{motta2020determining}.
The QITE method is based on the imaginary-time
Schr\"{o}dinger equation
\begin{equation}\label{eq:dynamic}
    \frac{d}{d\tau }\ket{\psi(\tau)}=-(H-E_\tau)\ket{\psi(\tau)},
\end{equation}
where $E_\tau=\langle\psi(\tau)|H| \psi(\tau)\rangle$ is the instantaneous expectation value of the Hamiltonian. 

The normalized quantum state at imaginary time $\tau$ is given by 
\begin{equation}
|\psi(\tau)\rangle:=\frac{e^{-H \tau}|\psi(0)\rangle} {\sqrt{\left\langle\psi(0)\left|e^{-2 H \tau}\right| \psi(0)\right\rangle}}.    
\end{equation}
The ground state of the system Hamiltonian $H$ can be obtained as the long-time limit of this evolution: $\lim _{\tau \rightarrow \infty} \frac{|\psi(\tau)\rangle}{\|\ket{\psi(\tau)}\|}$. 
However, a major limitation of QITE is that the required quantum circuit depth grows exponentially with the correlation domain size (approximately the system's correlation length) and linearly with the number of imaginary time steps~\cite{gomes2021adaptive}.

An alternative approach is to combine the QITE with variational quantum eigensolver (VQE), resulting in the Variational Quantum Imaginary-Time Evolution (VQITE) method~\cite{Yuan2019theoryofvariational,mcardle2019variational}. VQITE approximates the imaginary-time evolution within a fixed-depth variational ansatz by minimizing an energy-based cost function. It can also be interpreted as a special case of VQE employing quantum natural gradient optimization~\cite{stokes2020quantumnatural}.

Let us briefly review the fundamental framework of VQITE.
Consider a parameterized quantum state $\ket{\psi(\bm{\theta}(\tau))}$, where the real-valued parameters $\bm{\theta}$ evolve with the imaginary time $\tau$.
Substituting this ansatz into the dynamic Eq.~\eqref{eq:dynamic}  yields
\begin{equation}\label{eq:para_dynamic}
    \left[\sum_i \dot{\theta}_i\frac{\partial }{\partial \theta_i}+(H-E_\tau)\right]\ket{\psi(\bm{\theta})}=\bm{0}. 
\end{equation}

The evolution in imaginary time can then be approximated by updating the parameters according to: 
\begin{equation}
   \bm{\theta}(\tau +\delta \tau)\approx\bm{\theta}(\tau)+\delta\tau \dot{\bm{\theta}}.
\end{equation}
where $\delta \tau$ is a small time step. Hence, the key step in VQITE is determining the time derivative of the parameters $\dot{\bm{\theta}}$. 

Following McLachlan’s variational principle, which minimizes the norm of the residual of the imaginary-time Schr\"{o}dinger equation
\begin{equation}\label{eq:mclachlan}
    \delta \left |\left| (\frac{d}{d\tau} + H - E_\tau)\ket{\psi(\tau)}\right |\right|=0. 
\end{equation}
Under this principle, the evolution of the parameters leads to the following linear system~\cite{Yuan2019theoryofvariational} 
\begin{equation}\label{eq:evolution}
    A^R \dot{\bm{\theta}}=-C^R,
\end{equation}
where the real-valued matrices $A^R$ and $C^R$ are defined as:
\begin{align}
\left(A^R\right)_{i j} &=\operatorname{Re}\left[\frac{\partial\langle\psi(\boldsymbol{\theta}(\tau))|}{\partial \theta_i} \frac{\partial|\psi(\boldsymbol{\theta}(\tau))\rangle}{\partial \theta_j}\right],\\
\left(C^R\right)_i&=\operatorname{Re}\left[\frac{\partial\langle\psi(\boldsymbol{\theta}(\tau))|}{\partial \theta_i} H|\psi(\boldsymbol{\theta}(\tau))\rangle\right].
\end{align}
Solving this system yields the parameter update rule:
\begin{equation}
    \dot{\bm{\theta}}=-(A^R)^{-1} C^R.
\end{equation}
\end{appendix}

\end{document}